\documentclass[twocolumn]{aastex6}

\usepackage{xcolor}
\usepackage{amsmath}
\usepackage{affils}
\usepackage{booktabs}

\newcommand{\asamp}[0]{visual sample}
\newcommand{\ha}[0]{H$\alpha$}
\newcommand{\sbunits}[0]{mag arcsec$^{-2}$}

\newcommand{\nsampparent}[0]{8412}
\newcommand{\nsampfield}[0]{6875}
\newcommand{\nsampauto}[0]{226}
\newcommand{\nsampgold}{101}
\newcommand{\fsampauto}{3.29\%}
\newcommand{\fsampautolz}{6.1\%}
\newcommand{\fsampgold}{1.5\%}
\newcommand{\fsampgoldlz}{2.6\%}

\begin{document}
\shortauthors{Kado-Fong et al.}

\title{Star Formation in Isolated Dwarf Galaxies Hosting Tidal Debris:\\
Extending the Dwarf-Dwarf Merger Sequence}

\DeclareAffil{princeton} {Department of Astrophysical Sciences, Princeton University,
Princeton, NJ 08544, USA}
\DeclareAffil{carnegie} {The Observatories of the Carnegie Institution for Science, 813 Santa Barbara Street, Pasadena, CA 91101, USA}
\DeclareAffil{ohio}{Center for Cosmology and AstroParticle Physics (CCAPP), The Ohio State University, Columbus, OH 43210, USA}
\DeclareAffil{naoj}{National Astronomical Observatory of Japan, 2-21-1 Osawa, Mitaka, Tokyo 181-8588, Japan}
\DeclareAffil{sokendai}{Graduate University for Advanced Studies(SOKENDAI), 2-21-1 Osawa, Mitaka, Tokyo 181-8588, Japan}

\affilauthorlist{
  Erin Kado-Fong\affils{princeton},
  Jenny E. Greene\affils{princeton},
  Johnny P. Greco\affils{ohio},
  Rachael Beaton\affils{princeton,carnegie},
  Andy D. Goulding\affils{princeton},
  Sean D. Johnson\affils{princeton,carnegie},
  Yutaka Komiyama\affils{naoj,sokendai}
  }
  
  \date{\today}

\begin{abstract}
Like massive galaxies, dwarf galaxies are expected to undergo major mergers with other dwarfs. However, the end state of these mergers and the role that merging plays in regulating dwarf star formation is uncertain. Using imaging from the Hyper Suprime-Cam Subaru Strategic program, we construct a sample of dwarf-dwarf mergers and examine the star formation and host properties of the merging systems.
These galaxies are
selected via an automated detection algorithm from a sample of \nsampfield{} 
spectroscopically selected isolated dwarf galaxies at $z<0.12$ and 
$\log(M_\star/M_\odot)<9.6$ from the Galaxy and Mass
Assembly (GAMA) and Sloan Digital Sky Survey (SDSS) spectroscopic campaigns.
We find a total tidal feature detection fraction of \fsampauto{} (\fsampautolz{} when 
considering only galaxies at $z<0.05$). The 
tidal feature detection fraction rises strongly as a function of star
formation activity; 15\%-20\% of galaxies with extremely high \ha{} equivalent width
(\ha{} EW $>$ 250\AA) show signs of tidal debris. Galaxies that
host tidal debris are also systematically bluer
than the average
galaxy at fixed stellar mass. These findings extend the observed dwarf-dwarf
merger sequence with a significant sample of dwarf
galaxies, indicating that star formation triggered in mergers between
dwarf galaxies continues after coalescence.
\end{abstract}

\section{Introduction}
Dwarf galaxies are the most abundant subset of galaxies in the universe
\citep{binggeli1988}. Though hierarchical structure formation should
also proceed for these systems,
there are very few examples of extragalactic dwarf-dwarf mergers in the 
literature. A number of individual cases have been examined in great
detail
\citep{martinezdelgado2012, annibali2016, privon2017},
 and a small number of systematic searches for dwarfs undergoing
 interactions with other low-mass systems have been performed
 either by searching for dwarf companions \citep{stierwalt2017}, or by
 searching for low surface brightness (LSB) merger signatures
 \citep{paudel2018}.
 
Such studies show evidence for hierarchical merging at low galaxy
mass -- \cite{annibali2019} finds evidence for ongoing accretion events 
around the dwarf galaxy DD0 68 with mass ratios of
10:1 and 100:1. Though these focused studies reveal in great 
detail the accretion histories of individual systems, the question
of the commonality of merger features around dwarf galaxies 
is still poorly understood.

From matching SDSS to Millenium-II, $\approx3\%$ of dwarf galaxies 
($\log(M_\star/M_\odot)<9.6$) should have a companion 
of at least $\sim 30\%$ its mass \citep{sales2013, besla2018}. 
Moreover, $\sim10\%$ of dwarf galaxies and $15-20\%$ of dwarfs located
far from a massive neighbor are expected to have
undergone a major merger since $z\sim1$ \citep{deason2014}. 
It is thus of interest to 
characterize the frequency and characteristics of dwarf galaxies that show 
signs of a recent dwarf-dwarf merger.

The star formation activity during and after the merger of two dwarf 
galaxies is also a largely unconstrained realm. 
Less than 2\% of dwarf galaxies in the
field are observed to be quiescent \citep{geha2012}. If dwarf galaxies
are indeed undergoing major mergers, this suggests that mergers between
dwarfs do not succeed in quenching star formation, 
in contrast to the expectation of major merger outcomes for more
massive galaxies \citep{bekki1998, hopkins2008, ellison2018}. 

Furthermore, such dwarf-dwarf mergers have
been proposed as a mechanism to create Blue
Compact Dwarfs (hereafter BCDs). These
starbursting dwarfs are at $M_B \geq -18$, have physical sizes of
less than 1 kpc, and show a strongly starbursting spectrum \citep{bekki2008}.
BCDs tend to show a more centrally concentrated mass profile than a typical dwarf
irregular galaxy, as well as higher central surface brightnesses
\citep{janowiecki2014}. However, in deeper observations, BCDs  are also found to host old stellar
populations of $\gtrsim 1$ Gyr \citep{aloisi2007, annibali2013}.

Previous searches for tidal features around low mass galaxies have found
a depressed tidal feature detection fraction relative to samples of more
massive galaxies.
A study of SDSS and CFHT Legacy Survey data 
found that approximately 0.68\% of the low mass galaxies in their
sample host tidal features \citep{paudel2018}. 
Though the tidal feature fraction should increase 
as the effective surface brightness limit 
improves \citep[see, e.g.][]{annibali2019}, 
it is possible that tidal features around
dwarf galaxies at low-$z$ are intrinsically rare features. Thus, in
this work we search for LSB tidal debris around
a sample of isolated dwarf galaxies identified in the GAMA and
SDSS spectroscopic surveys at $z<0.12$
in imaging from the Wide layer of the 
Hyper Suprime-Cam Subaru Strategic Program 
\citep[HSC-SSP;][]{HSC1stDR,HSC1styrOverview,Miyazaki18HSC,komiyama2018,Kawanomoto18HSC,Furusawa18HSC,Bosch18HSC,Haung18HSC,Coupon18HSC}.  

In \autoref{selecsec} we establish the parent sample of spectroscopically
confirmed dwarfs with imaging in HSC-SSP. In \autoref{detsec} we 
outline the automated detection algorithm employed to search for tidal
debris around dwarf galaxies. We present the dependence of the frequency
of detectable features around a host galaxy as a function of the host
properties and star formation activity in \autoref{ressec}. Finally,
we discuss the implications for the observational dwarf galaxy merging
sequence in \autoref{discusssec}.

Throughout this paper we adopt a standard flat $\Lambda$CDM model 
in which H$_0=70$ km s$^{-1}$ Mpc$^{-1}$ and $\Omega_m=0.3$.

\section{Sample Selection}\label{selecsec}
For the purposes of this study, we follow \citet{stierwalt2015} in defining a dwarf galaxy
as a galaxy with stellar mass of $\log(M_\star/M_\odot) < 9.6$. The stellar masses reported here are for individual dwarfs. We find only one case in which both galaxies in an interacting pair host tidal features. The dwarfs are sufficiently separated such that their individual stellar masses may be estimated.
We note that this is somewhat lower than the value used 
by other studies (for example,
\citet{paudel2018} uses a cutoff of $\log(M_\star/M_\odot) < 10.0$ in combined stellar mass for the dwarf and its companion, if applicable). 

We additionally exclude galaxies at $z>0.13$, as we are in practice unable to 
find LSB tidal debris around galaxies at this mass range above this redshift 
cutoff (for a discussion of the completeness of our tidal feature detection method,
see \autoref{zmass_fdet}). 

\subsection{HSC-SSP Imaging}
The detection of tidal debris resulting from low mass mergers requires imaging that 
covers a large enough area to find a sizable number of galaxies that host such features
while simultaneously probing the relevant low surface brightness universe. 
HSC-SSP is particularly well-suited towards this aim; HSC-SSP will cover over
1400 square degrees on the sky to a limiting magnitude of i$_{\rm HSC}\sim 26$~mag for point sources
\citep{bosch2018}. 
Detection of low surface brightness features is a function of both the detection algorithm and
the imaging sensitivity. 
We find that HSC-SSP reaches surface brightness limit of $\sim$27 mag arcsec$^{-2}$ when
detecting isolated LSB structure \citep{greco2018, kadofong2018}.  
When the morphology of the LSB structure is known (e.g. for measurements of
smooth stellar halos), measurements of individual galaxies reach
$>28.5$ mag arcsec$^{-2}$ \citep[massive ellipticals,][]{huang2018},
and measurements on stacked images reach $>30$ mag arcsec$^{-2}$ 
\citep[isolated central galaxies, on the order of one thousand objects stacked,][]{wang2019}.

For this work, we use the HSC S18A data release, which covers over 300 square
degrees on the sky in $g_{\rm HSC}$, $r_{\rm HSC}$, and $i_{\rm HSC}$. 
We do not use $z_{\rm HSC}$ and $y_{\rm HSC}$ imaging because the surface brightness limits and seeing
of these bands are significantly worse than the bluer bands, and including these bands would decrease the 
usable area of this work without contributing significantly to the detection of LSB structure.
This data release is equivalent to the second public data release
(hereafter PDR2) presented in \citet{aihara2019}. The area covered by our search is somewhat
smaller than the total area released in PDR2, as we require imaging in $g_{\rm HSC}$, $r_{\rm HSC}$, and $i_{\rm HSC}$ bands for
each target galaxy. We also briefly note that the background subtraction method implemented
in this release has been updated from that which was used for the first HSC-SSP public data release.
This new background subtraction method reduces oversubtraction of the halos around bright
galaxies, allowing us to target galaxies at lower redshifts than were possible in \cite{kadofong2018}.

\subsection{Spectroscopic sample}\label{specsample}
To generate a sample of isolated dwarfs, we consider 
only galaxies with spectra from either
the Sloan Digital Sky Survey (SDSS) spectroscopic surveys 
\citep[both legacy and BOSS surveys,][]{strauss2002, dawson2013, reid2016} or
from the Galaxy and Mass Assembly (GAMA) spectroscopic survey \citep{baldry2018}.

These cuts leave a parent sample of \nsampparent{} 
dwarf galaxies. 3733 of the target galaxies
have spectra from the SDSS surveys, while 5001 galaxies have spectra from GAMA
(678 galaxies have spectra from both GAMA and SDSS).
For galaxies with GAMA spectra, we adopt the stellar masses provided by the GAMA team \citep{taylor2011}.
These stellar masses are measured assuming a Chabrier initial mass function \citep{chabrier2003}. For 
galaxies with SDSS spectra, we adopt stellar masses derived using the \cite{conroy2009} 
flexible stellar population synthesis (FSPS) models assuming a 
Kroupa initial mass function \citep{kroupa2001}. 
For galaxies in our sample that have spectroscopy from
both GAMA and SDSS, we find that the FSPS stellar masses 
are higher than the equivalent measurement in the 
GAMA catalog by a median of 0.08 dex and a median absolute deviation of 0.35 dex.
To reconcile this systematic shift, we reduce the masses derived from SDSS observations
by 0.08 dex, but note that including or excluding this shift does not affect
the results presented in this work. 

Because we want to study the star formation properties of these dwarfs, we 
also use the \ha{} line measurements provided by the GAMA and SDSS spectroscopic
databases. 

\subsection{Isolated dwarf sample}
Finally, because we are interested only in interactions between two dwarf galaxies,
we require that the target galaxies have a 3D physical separation of at least 1 Mpc 
from the nearest massive
galaxy ($\log(M_\star/M_\odot) > 10.0$) in the 
NASA Sloan Atlas \citep{blanton2011}, 
as measured from the comoving distances using the spectroscopic redshifts
of the target catalog and the NASA-Sloan Atlas (hereafter NSA). In this work, we use NSA version 1.0.1, which
was released with SDSS DR13 and reaches $z\sim0.15$.
 
Though the NSA is nominally complete only to $z=0.055$, we note that the fraction of
galaxies that are flagged as satellites is independent of redshift. To confirm that
we are not misclassifying satellite dwarfs, 
we compare the satellite fraction as function of
redshift for the sample as constructed above and the satellite fraction of the dwarfs with GAMA spectroscopy when matched to the 
GAMA spectroscopic catalog, which is deeper than the NSA. We find no significant differences in the satellite 
fraction, and therefore 
conclude that we are not misclassifying a significant number of satellite galaxies
as field dwarfs.

We also remove galaxies that are projected closer than 0.01 deg 
to a massive galaxy, regardless of physical association. This cut aims to
remove those cases in which a dwarf galaxy overlaps significantly with 
high surface brightness light from a more massive projected neighbor. 

The isolation criterion that we adopt here is 
somewhat different than isolation criteria adopted in the literature in that
we make a cut in comoving distance rather than directly in velocity
space \citep{geha2012, paudel2018}. Because the isolation of the galaxies in 
question is important to the interpretation of this work, we additionally
verify that the choice of isolation criterion does not affect our results; we
refer the reader to \autoref{appendixB} for a discussion therein.

\begin{figure}[ht!]
\center{\includegraphics[width=\linewidth]{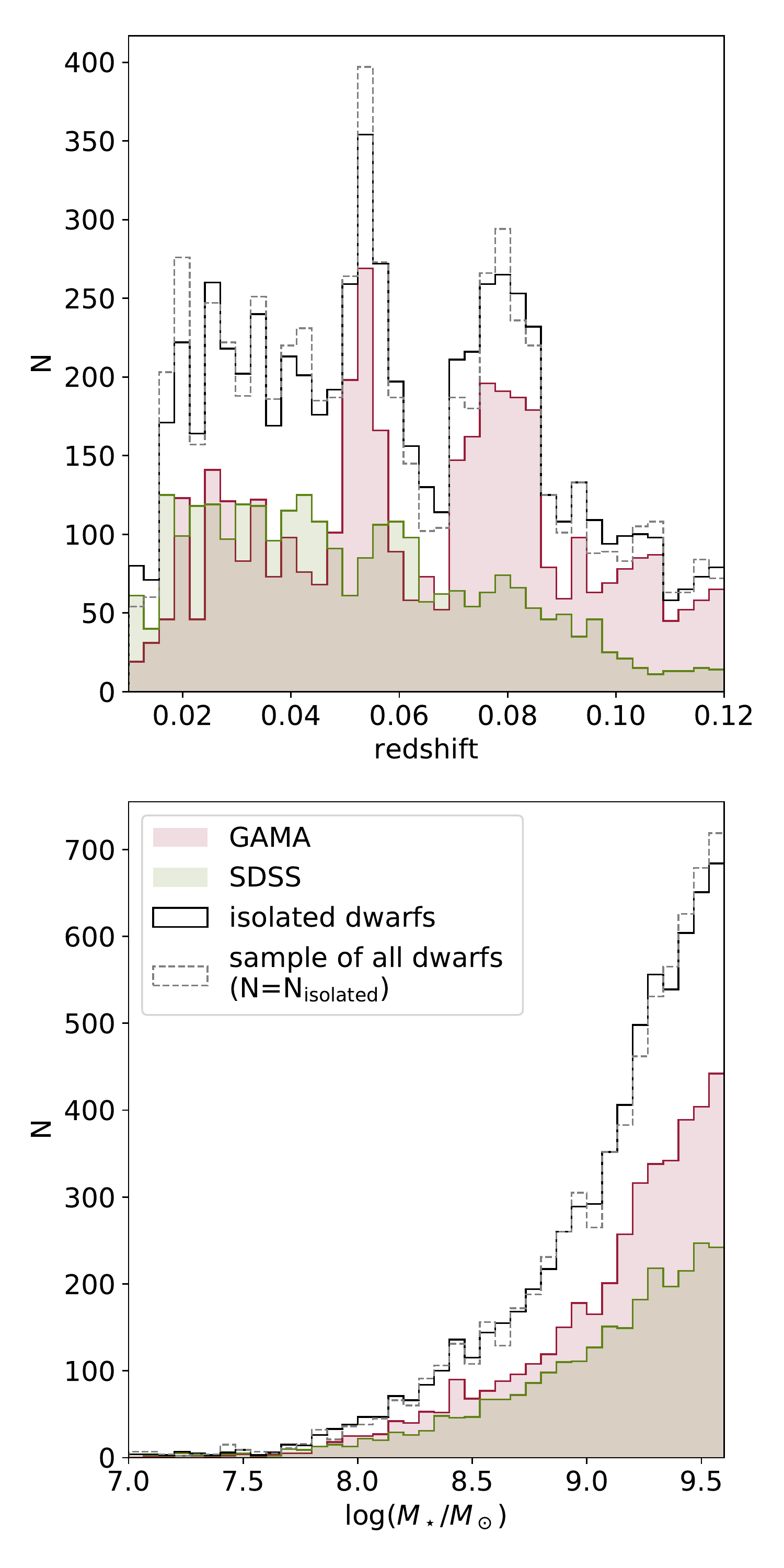}}
\caption{
The distribution of our sample in redshift (top panel) and stellar mass (bottom panel). In both panels,
the red filled histogram shows the GAMA dwarfs and the green filled histogram shows the SDSS 
dwarfs. The unfilled black and dashed grey histograms show the distribution of
the full isolated dwarf sample and a random subset of \nsampfield{} galaxies from the
full dwarf sample. This is to guide the eye and show that
show that there is no significant difference between the distributions 
of the full and isolated dwarf samples.
}
\label{parentsample_statistics}
\vspace{20pt}
\end{figure}

These distance cuts leave a final sample of \nsampfield{} galaxies; 
3169 of these galaxies
have spectra from SDSS, and 4034 have spectra from GAMA. Of these, 520 galaxies have spectra
from both SDSS and GAMA. In \autoref{parentsample_statistics}, we show the
distribution of the GAMA and SDSS dwarfs as a function of redshift (top) and
stellar mass (bottom). We also show the distribution of the isolated dwarf 
sample in black. To show that the isolation cut does not introduce a shift in
the redshift or stellar mass of the parent sample, we also show a random subset
of \nsampfield{} galaxies from the full dwarf sample as the dashed grey histogram.

Though the distribution of the SDSS and GAMA galaxy samples are markedly 
different in redshift and stellar mass, we will show that we reach the 
same conclusions when considering only the SDSS or only the GAMA sample. In cases where
spectra from both SDSS and GAMA are available,
we prioritize measurements from GAMA because the spectra are deeper. This 
will be discussed more fully in \autoref{detsec} and \autoref{appendixB}.

\begin{figure*}[ht!]
\center{\includegraphics[width=\linewidth]{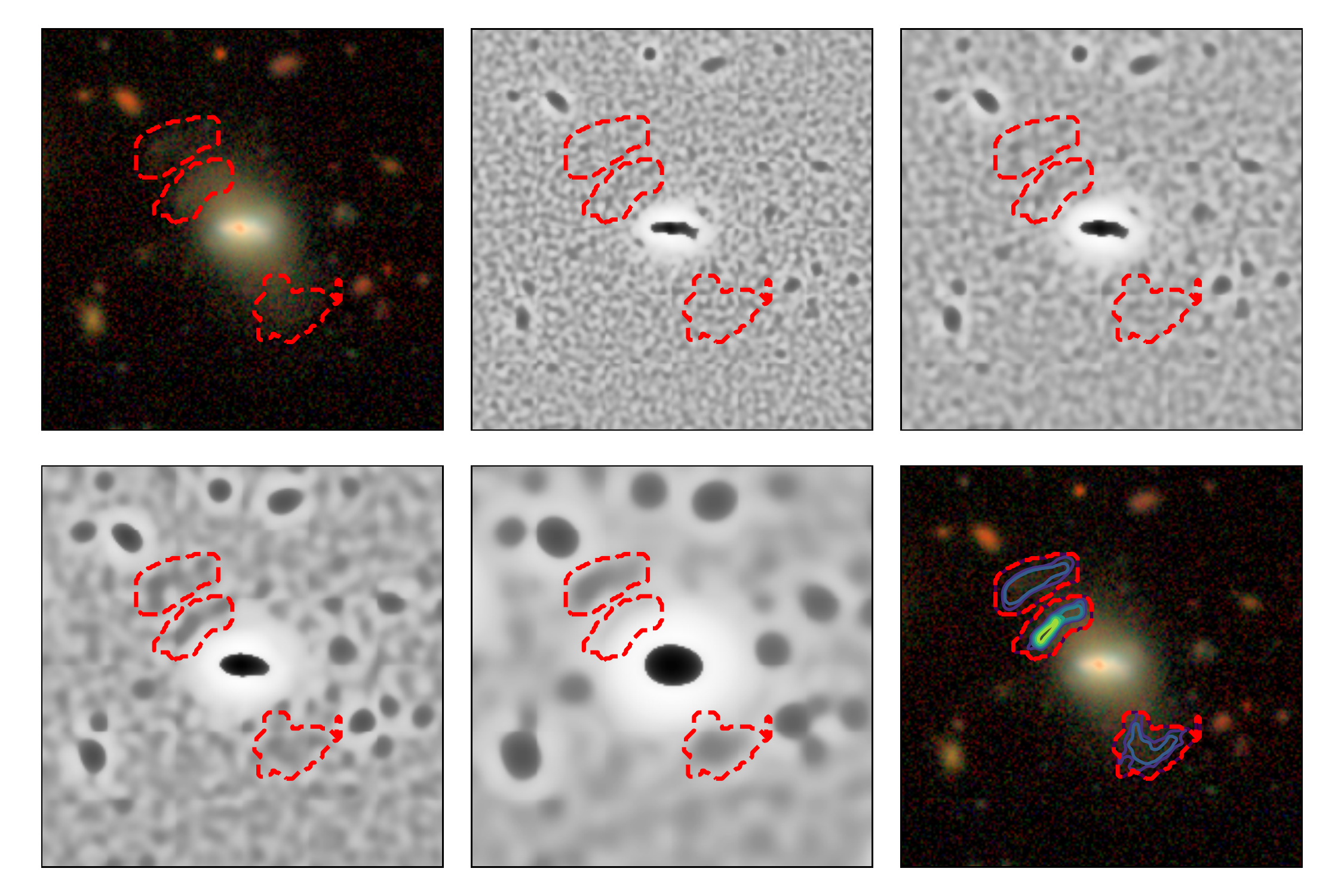}}
\caption{
An illustration of the multiscale detection method used to find tidal debris 
around dwarf galaxies. \textit{Top left:} the $gri$ composite image of the 
target galaxy. To guide the eye, the maximal area contour of the detection
map is shown 
as a contour in red in all panels. \textit{From top middle:} the starlet 
coefficients in order of increasing spatial scale. The intensity of the images
are scaled logarithmically. We note that 
certain tidal features are only seen in certain coefficients. For example, the 
inner shell is strongly detected in the second and third coefficients, while
the outer shells are detected with high significance in the third and fourth 
coefficients. \textit{Bottom right:} the $gri_{\rm HSC}$ composite image of the target 
galaxy with the full detection map overlaid as a contour. The contours show
regions of increasing detection significance (i.e. detected in more coefficients).
}
\label{starlet}
\vspace{20pt}
\end{figure*}

\begin{figure*}[ht!]
\center{\includegraphics[width=\linewidth]{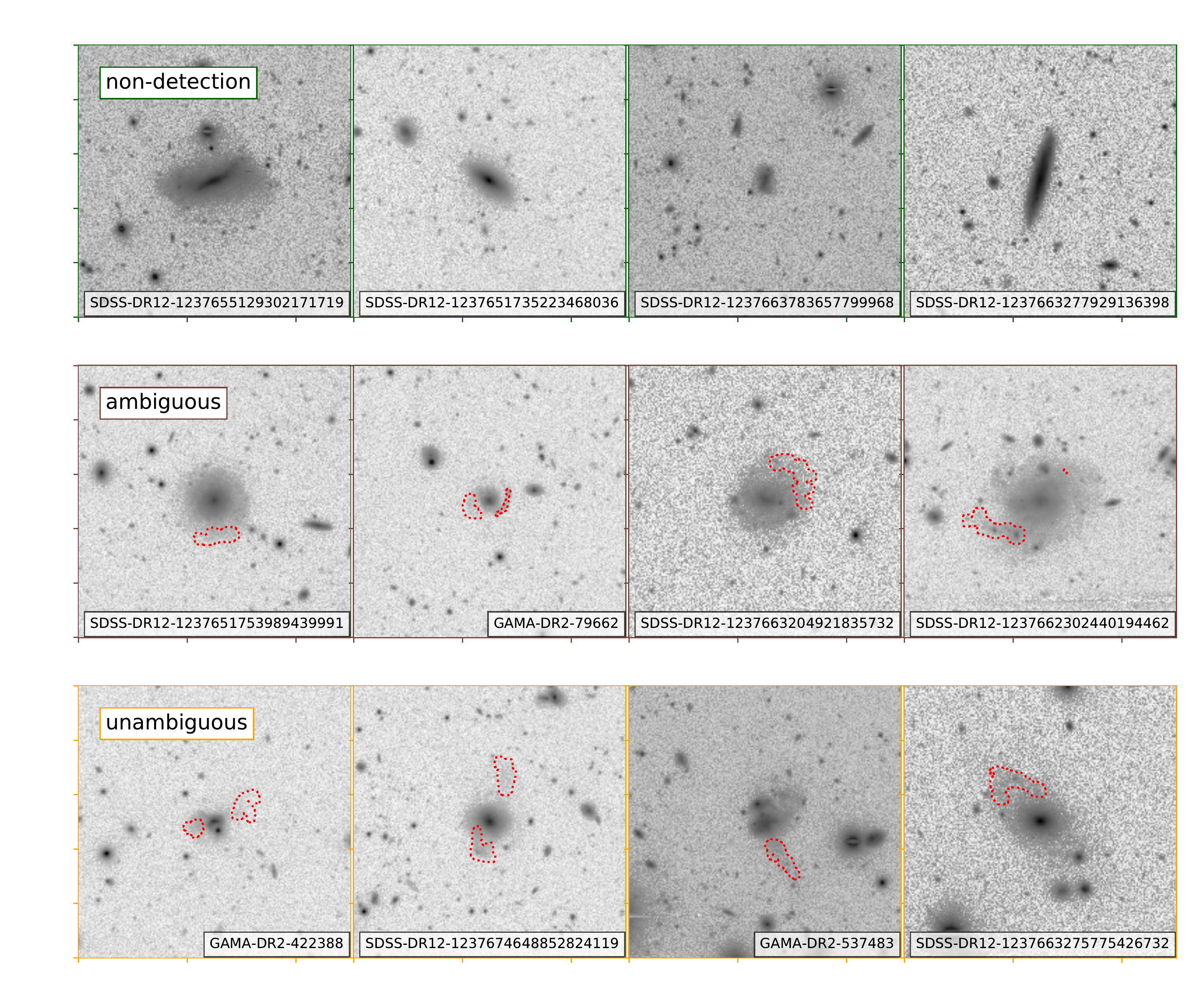}}
\caption{
Examples of galaxies with no tidal feature detection (top row), ambiguous tidal
feature detections (middle row), and unambiguous tidal feature detections (bottom
row). The HSC spectroscopic redshift catalog ID is shown in the bottom right of each panel. For the examples in which candidate tidal feature systems are detected,
the lowest significance level detection map is plotted as a red outline. We note
that although there are some clear cases of spiral arm contamination (e.g. 
the far right middle panel), we do not remove these cases from the sample, and show 
that they do not statistically affect our conclusions.
}
\label{examples}
\vspace{20pt}
\end{figure*}

\begin{figure*}[ht!]
\center{\includegraphics[width=\linewidth]{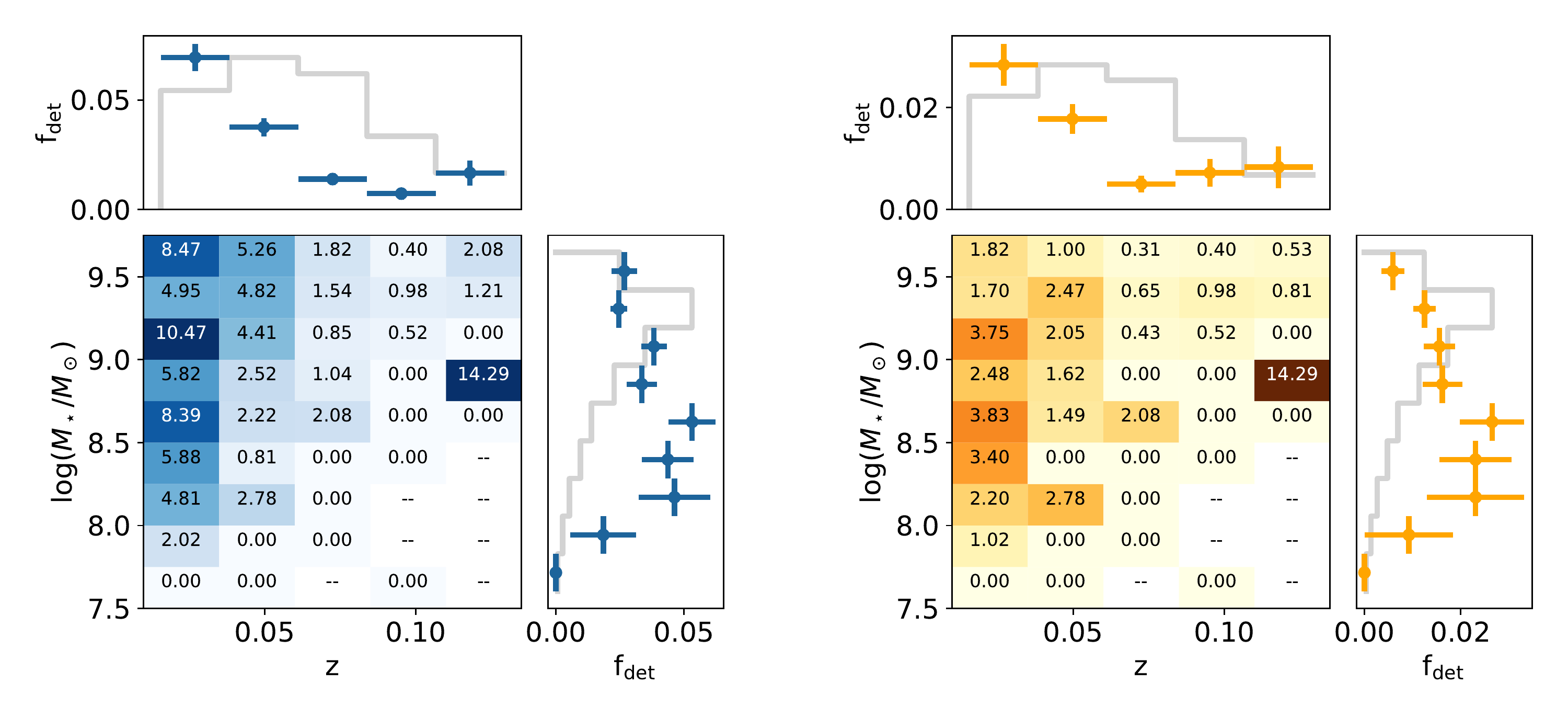}}
\caption{
Tidal feature detection percentage for our sample of dwarfs in bins of redshift and
host stellar mass (main panel). The color and numbers reflect the
percentage of dwarf galaxies that host tidal features in each bin. 
The top auxiliary panels show the projection of the detection fraction
distribution onto redshift, while the right-hand auxiliary panels show the projection as a
function of host stellar mass. In each auxiliary panel, the unfilled grey histogram shows the
normalized distribution of the parent sample. The left panels in blue show the automated sample, while
the right panels in orange show the \asamp{}.
}
\label{zmass_fdet}
\vspace{20pt}
\end{figure*}
\section{Tidal Feature Detection}\label{detsec}
From previous work, the fraction of dwarf galaxies that 
host tidal features detectable in HSC-SSP is likely on the order of a few percent \citep{paudel2018}; 
though it is possible to 
select tidal feature hosts via visual inspection, constructing an automatic detection algorithm is a significantly more scalable 
approach. Towards this end, we use an updated version of the tidal feature detection algorithm presented in 
\cite{kadofong2018} to automatically identify tidal features around dwarf galaxies. These tidal feature hosts are selected from the sample of \nsampfield{} field dwarfs identified in \autoref{specsample}.

\subsection{Updates to the method of \cite{kadofong2018}}
The main purpose of the algorithm presented in \cite{kadofong2018} is to detect 
tidal features against the smooth background of a host galaxy by leveraging the 
contrasting spatial scales of the tidal feature and the host halo. To do so in a way that
is independent of the morphology of the host, we decompose the image into coefficients
of increasing spatial scale. Tidal debris candidates are then identified as contiguous structures
in the spatial decomposition.

Here, we summarize the updated detection algorithm. 
For the full details of the updates made to the method, see \autoref{appendix}.

In order to detect tidal features on a variety of spatial scales, we decompose the image of the 
target galaxy into wavelet coefficients of increasing characteristic length; specifically, we use the 
starlet transform of \cite{starck2015}. Low surface brightness 
features are detected in each coefficient independently and assigned to the most probable host galaxy
(as determined by flux-weighted distance). Tidal features are detected independently in the 
g$_{\rm HSC}$, r$_{\rm HSC}$, and i$_{\rm HSC}$ band images; tidal features must be detected in 
at least two bands in order to be accepted as part of the final detection map.

In \autoref{starlet} we illustrate the detection method of a galaxy in our sample.
The RGB images show the $gri$ composite image of the galaxy, while the grey scale
images show the wavelet coefficients (the first wavelet coefficient is not shown)
of the $i_{\rm HSC}$ band image. The full detection map is shown in the bottom right panel
of the figure.

We apply this algorithm to the sample of \nsampfield{} field dwarfs, and identify \nsampauto{} dwarfs with detectable low surface brightness debris.

\subsection{Construction of the \asamp{}}
The main purpose of the algorithm described above is to avoid the need to 
visually classify the full sample. However, because
the morphology of many dwarf galaxies are irregular, 
there exist ambiguous cases in which the origin of a possible tidal feature
is unclear. To ensure that the inclusion of such 
ambiguous cases does not bias our results, we visually inspect the
automated sample and construct a  
``\asamp{}'' subset that consists of only tidal feature systems that are 
unambiguously the result of a merger with another dwarf galaxy. We follow the 
benchmarks set forward in \cite{paudel2018}, and generally look for the morphology
of the detected LSB feature to be inconsistent with extended irregular structure or 
flocculent spiral arms.

We find that \nsampgold{} galaxies out of the \nsampauto{} in the automated sample host unambiguous signs of a dwarf-dwarf merger, and are included in the visual sample. We find four general classes of rejected objects: tidal arms/stirring, amorphous and asymmetric LSB structure, false detections from overlapping sources, and false detections from imaging artifacts. Though the existence of extended tidal arms is a signpost of interaction, such features can be formed both from an interaction with an equal mass companion \citep{toomre1972} and during an interaction with a more massive galaxy \citep[see, e.g.][]{villalobos2012,paudel2014,hendel2015}. Because the dwarf galaxies are selected to be isolated, we expect that the majority of these cases are indeed due to dwarf-dwarf mergers, but there may be some exceptions. Tidal arms account for 50\% of the ambiguous galaxies. The formation mechanism behind individual amorphous and asymmetric LSB structures is unclear, and may be due to secular mechanisms; such cases account for 39\% of the ambiguous sample. Finally, false detections due to overlapping sources (e.g. overlapping tidal features, galactic cirrus) and false detections due to imaging artifacts account for 6\% and 5\% of the ambiguous sample, respectively. False detections thus account for 6\% of the total automated sample. Because some of the ambiguous cases are consistent with being formed via an interaction with a more massive galaxy, here we re-emphasize that our choice of isolation criterion does not affect our results (see \autoref{appendix}).

\autoref{examples} shows a selection of example galaxies with no tidal feature
detection (top row, green outlines), with an ambiguous detection (middle row, 
brown outlines), and with an unambiguous detection (bottom row, orange outlines). 

We will use this \asamp{} to demonstrate that when considering trends with 
respect to the presence of tidal features, 
our results are unchanged whether or not 
ambiguous cases are included. 
When considering an individual galaxy or when asking a question that
requires tidal features that are photometric quality, 
we recommend using the \asamp{} to determine whether there are 
unambiguous signs of merging between dwarf galaxies. 
As we will show below, however,
we find no statistical differences in the population properties of interest between
the automated sample and \asamp{}. In \autoref{summarystats} we provide a summary
of the total, field, and interacting dwarf samples.

We publish our catalog of dwarf-dwarf mergers (and non-detections) 
in machine-readable format along with this work. 
For the reader's convenience, we include stellar mass estimates, \ha{} flux
and equivalent width measurements, and $(g-i)$ colors in the catalog. A sample of this
table is given in \autoref{mrt}.

\begin{table}
\begin{center}
\begin{tabular}{lcc}
\toprule
{} &    N & N$_{\rm detection}$  \\
\midrule{}
all dwarfs            &  \nsampparent{} & 284  \\
field dwarfs          &  \nsampfield{} & \nsampauto{}  \\
\asamp{} &  -- & \nsampgold{}        \\
\bottomrule
\end{tabular}
\caption{Summary of the merging dwarf sample presented in this work. The first
row gives the total number of dwarfs and number of tidal feature hosts for all dwarfs
with spectroscopic confirmation and HSC imaging (i.e. including dwarfs that are 
non-isolated). The second row gives the same numbers, but for only those dwarfs that
satisfy the isolation criterion described in \autoref{specsample}. 
The final row gives the number of tidal feature hosts whose LSB features are 
unambiguously the result of a merger with a companion.}
\end{center}
\end{table}\label{summarystats}

\begin{figure*}[ht!]
\center{\includegraphics[width=\linewidth]{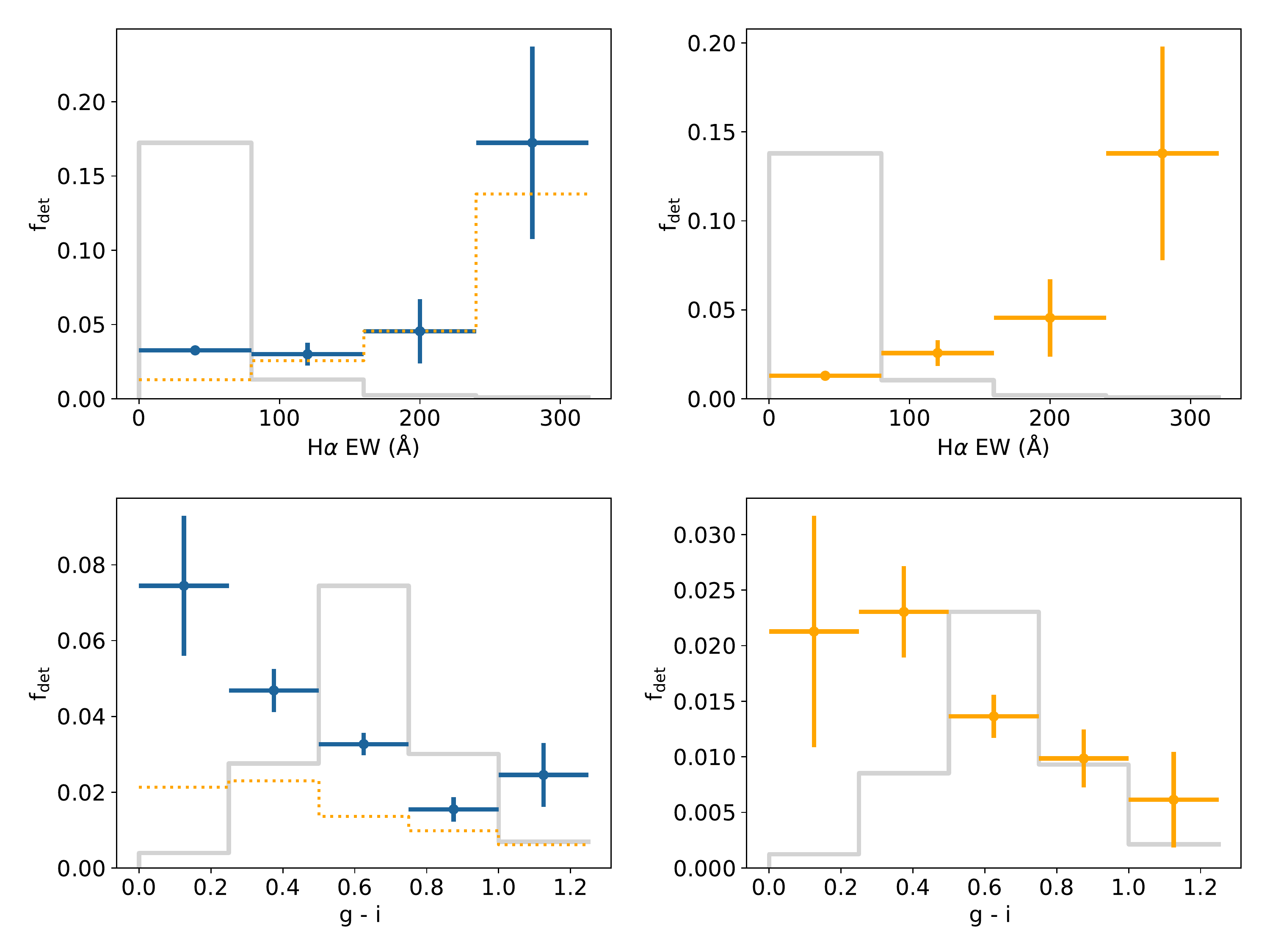}}
\caption{
The fraction of dwarfs around which tidal debris is detected for the automated sample (left)
and \asamp{} (right). To guide the eye, the detection fraction for the \asamp{} is 
also plotted as a dotted gold line in the left column panels.
The top row shows detection fraction as a function of \ha{} equivalent width,
while the bottom row shows detection fraction as a function of host $(g-i)$ color as measured
by SDSS. In each panel, the unfilled grey histogram shows the normalized distribution
of the parent sample. 
}
\label{sfrcolor_fdet}
\vspace{20pt}
\end{figure*}

\section{Results}\label{ressec}
In total, our automated sample consists of \nsampauto{} galaxies; \nsampgold{} of these
galaxies make up the \asamp{} (i.e. are visually confirmed to be unambiguous 
merger debris). 

We therefore find a total tidal feature detection fraction of \fsampauto{}
in the automated sample and \fsampgold{} in the \asamp{}. When considering only 
those host galaxies at $z<0.05$, we find a tidal feature detection fraction of
\fsampautolz{} in the automated sample and \fsampgoldlz{} in the \asamp{}. We consider
a more detailed analysis of our completeness as a function of host stellar mass
and redshift in \autoref{completeness}.

\subsection{Completeness of the automated and visual samples}\label{completeness}
In \autoref{zmass_fdet} and \autoref{sfrcolor_fdet} we show the statistical
properties of the automated and \asamp{} in blue and orange, respectively. 
In particular, we note that although the absolute fraction of galaxies with tidal
features is shifted down for the \asamp{}, there is no significant shift in the
relative tidal feature detection fraction as a function of any of the 
host galaxy properties that we examine.

In particular, \autoref{zmass_fdet} shows the detection fraction for the 
automated sample and \asamp{} as a function of host galaxy mass and redshift.
The main panels show the tidal feature detection fraction at a given redshift and
host stellar mass, while the framing panels show the detection fraction projected along
stellar mass (right) and redshift (top). 

Both when considering host stellar mass and host redshift, we see that the 
distribution of host galaxy properties does not change significantly between
the automated sample and the \asamp{}. In particular, we find that the 
redshift of 50\% relative completeness is approximately $z=0.05$ for both samples.
Our absolute completeness is unknown, though our previous study at higher masses indicated that this algorithm is sensitive to tidal debris down to $\sim 27$ \sbunits{} and decreases in completeness for very bright ($\lesssim 24.5$ \sbunits{}) features \citep{kadofong2018}. 
Nevertheless, we can measure our relative
completeness by comparing to the tidal feature detection fraction 
at the low redshift end of our sample (\fsampautolz{} and \fsampgoldlz{} at $z<0.05$)
in our sample and making the assumption that there is no astrophysical change in
true tidal feature occurrence fraction within the redshift range considered.

Similarly, we see that at the lowest redshift bin, where we are most complete, the
tidal feature detection fraction decreases as host stellar mass decreases at stellar masses $\log(M_\star/M_\odot) \lesssim 8.5$). 
This can
be understood by noting that as the host stellar mass decreases, the stellar surface 
density of tidal debris generated from a merger at a given mass ratio decreases.

Because the difference in the distribution of tidal feature detection fractions for the
automated and \asamp{} is not statistically significant, we conclude that 
astrophysical contamination by features that are not related to recent accretion events
 does not significantly affect the outcome of our results. We furthermore conclude that
some of the visually ambiguous features in the sample are indeed tidal in nature.

\subsection{Detection fraction and star formation}
In order to probe the effect of host star formation rate on the tidal feature incidence rate, 
we show the tidal feature detection fraction as a function of host \ha{} equivalent width and 
of host (g - i) color from SDSS in 
\autoref{sfrcolor_fdet}. Here, we use host colors from SDSS because the centers of
a subset of the brighter galaxies in our sample are saturated in HSC imaging.
Again, we find no significant difference between the
distribution of the automated and \asamp{} when considering tracers of star 
formation.
 
We find that the tidal feature detection fraction decreases monotonically 
as a function of host (g - i) color, with the detection
fraction of the bluest host galaxies ($(g-i)<0.3$) 
a factor of $\sim4$ higher than that of the reddest hosts ($(g-i)>0.9$) for both
the automated and \asamp{}.

To show that star formation rate does indeed drive this color dependence, 
we also show the tidal feature detection
fraction as a function of \ha{} equivalent width. At this mass range, almost the entirety ($>99\%$) of galaxies accessible to the SDSS and GAMA spectroscopic surveys are star forming with ${\rm EW}_{\rm H\alpha} > 0$.
As can be seen in \autoref{sfrcolor_fdet}, we find that the fraction 
of galaxies that host detectable tidal features increases as 
a function of \ha{} equivalent width.  

\autoref{sfrcolor_fdet} shows the results when considering the galaxies 
observed in the GAMA and SDSS spectroscopic surveys together. To confirm that
the dependence of tidal feature detection fraction on SFR is not a 
manifestation of the nature of the parent sample, in \autoref{appendixB} we show
that we find the same result when considering the GAMA and SDSS galaxies
separately.

If we define the starbursting 
sample as those hosts with \ha{} equivalent widths greater
than 100\AA{} \citep{lee2009, stierwalt2015}, 
we find that 4.8\% of the starbursts host 
detectable tidal features. In the highest \ha{} equivalent width bin
(240$<$EW$_{\rm H\alpha}<$320\AA), the fraction of 
galaxies that host tidal features grows to 17\%, albeit with large 
uncertainty as there are only 29 dwarfs in this bin, 5 of which show signs of
a merger.

\begin{figure}[ht!]
\center{\includegraphics[width=\linewidth]{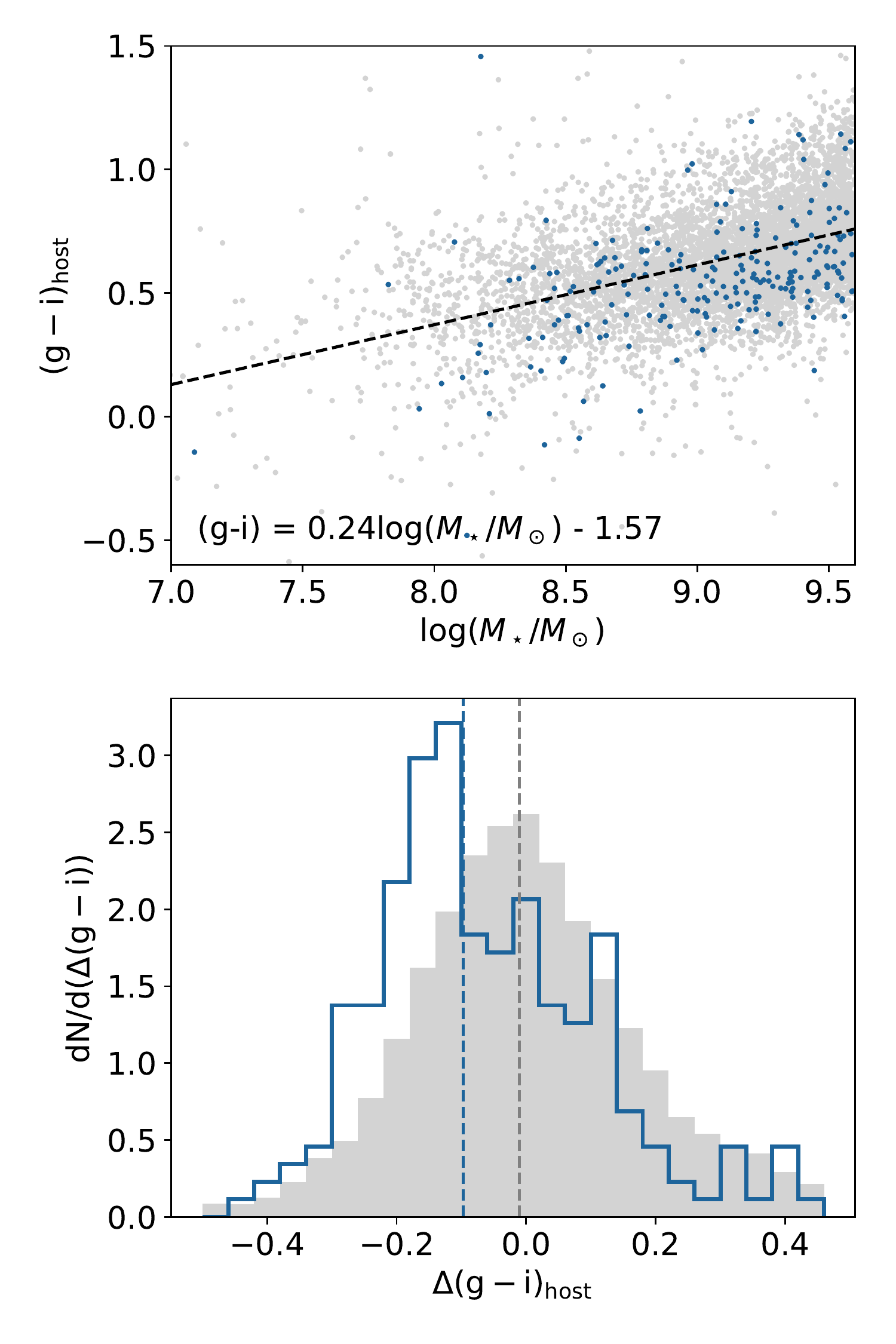}}
\caption{
\textit{Top:} $(g-i)$ color from SDSS versus stellar mass for non-interacting field dwarfs
(grey) and tidal feature host centers (blue). The dashed black line is fit to
the non-interacting dwarf galaxy sample. \textit{Bottom:} the deviation of host
$(\rm g-i)$ color from the
linear fit shown in the top panel for the non-interacting hosts (grey histogram) and tidal 
feature hosts (blue unfilled histogram). The dashed grey and blue vertical lines show the
median deviation for the non-interacting and tidal feature hosts, respectively.
}
\label{deltagminusi}
\vspace{20pt}
\end{figure}

\subsection{Host color vs. stellar mass}
In the top panel of \autoref{deltagminusi}, we show $(g-i)$ as a function of host
galaxy mass for the tidal feature hosts in blue
and the centers of apparently non-interacting field dwarfs (selected using
the same nearest neighbor cut as for the tidal feature hosts) in grey. The top panel 
scatter points show values for individual systems, while the bottom panel shows the
deviation of the central color from a fit to the $(\rm g-i)$ colors of the 
non-interacting hosts.

This illustrates that for their stellar mass, the central SDSS color of the \asamp{} 
tidal features hosts are
slightly bluer than the isolated dwarfs in which 
we detect no tidal features, with the median $\Delta (\rm g-i)$ values differing by
$\mathcal{S} = \langle \Delta (\rm g-i)_{\rm host, TF}\rangle_{50} - \langle \Delta (\rm g-i)_{\rm host, noTF} \rangle_{50} = -0.086$
(where $\langle \rangle_{50}$ refers to the median).

Because this analysis is sensitive to host stellar mass, we verify that the results do
not change under the following modifications: neglecting the shift in stellar mass
applied to the SDSS-derived quantities in \autoref{specsample}
($\mathcal{S} = -0.086$),
using only masses from SDSS measurements ($\mathcal{S} = -0.097$), 
and using only masses from GAMA measurements ($\mathcal{S} = -0.082$). In all cases, the
tidal feature hosts remain bluer than their non-interacting counterparts and the 
deviation from the fiducial value of $\mathcal{S}$ is small 
($\max (|\delta \mathcal{S} / \mathcal{S}|) = 0.12$). We repeat this 
analysis using \ha{} derived star formation rates of the $z<0.05$ SDSS 
galaxies, and find a statistically
significant shift ($p<0.001$) with a median increase in SFR of 0.07 dex
for the tidal feature hosts. We do not correct for reddening, as the measured Balmer decrements of the sample imply a negligible correction.

 \subsection{Tidal Feature and Host Morphology}
Although we do not visually classify the morphology in the full sample, 
we make a brief points on the tidal feature 
morphology present in the sample.

In particular, we note that there exists a significant population of 
dwarf galaxies that host stellar shells, similar to those observed in
higher mass galaxies \citep{atkinson2013,carlsten2017,hood2018,kadofong2018}. Such shells have also been observed around early-type dwarf galaxies in the Virgo Cluster \citep{paudel2017}.

In the literature, the formation of shells around galaxies is centered 
largely on the formation of shells around massive ellipticals. 
However, as observed in this sample, shells can also be formed around low-mass galaxies. 

\begin{figure*}[ht!]
\center{\includegraphics[width=\linewidth]{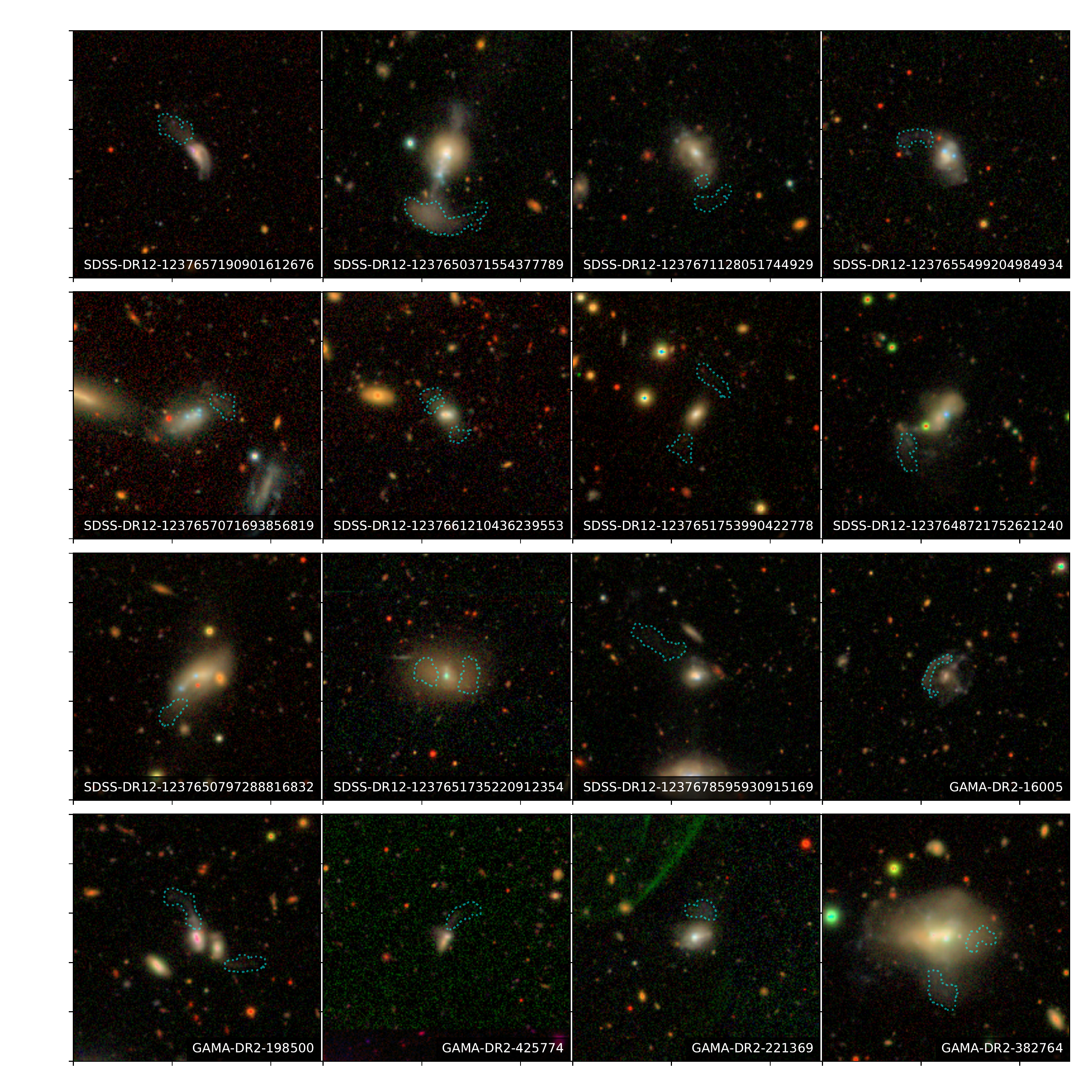}}
\caption{
$gri$-composite images for 16 of the starbursting dwarfs with detected tidal debris in our sample. The dashed cyan contours show the maximal outline of the tidal feature detection map produced from our algorithm. The HSC spectroscopic redshift catalog ID is shown in the bottom right of each panel.
}
\label{starbursts}
\vspace{20pt}
\end{figure*}

We also note that the starbursting galaxies in our sample that host 
tidal features have morphologies consistent with those of blue compact
dwarfs (hereafter BCDs). \autoref{starbursts} shows \textit{gri} composite
images for the set of starbursting dwarfs with detected tidal features in the sample.

\section{Discussion}\label{discusssec}


\subsection{The dwarf-dwarf merger sequence}
The observed increase in the fraction of galaxies that 
host tidal features as a function of \ha{} equivalent width 
(see \autoref{sfrcolor_fdet}) is consistent with the
picture in which some fraction of starbursting dwarfs 
are undergoing a period of triggered star formation as the 
result of a merger with another dwarf
galaxy.

We also find that dwarfs which host tidal features are systematically bluer
than those that do not host detectable tidal features, as shown in \autoref{deltagminusi}.
This offset can be interpreted either as an increase in the star formation 
efficiency (here we define the star formation efficiency to be 
${\rm SFE} = {\rm SFR}/M_{\rm H_2}$) or the molecular gas fraction 
($f_{\rm H_2}=M_{\rm H_2}/M_\star$)
in the post-merger galaxy (i.e. a movement down and to the right in
$(\rm g-i)$-$\log_{10}(M_\star/M_\odot)$ space), or as an increase in stellar 
mass while maintaining the same specific star formation rate (sSFR$={\rm SFR}/M_\star$, 
such that sSFR=$f_{\rm H_2}$SFE) and central host color (i.e.
a shift purely rightwards). We find that the median offset in host color can nominally be
reproduced by assuming that all mergers are the result of an equal mass merger with
no net change in the central host color. However, this interpretation cannot explain the 
increase in the tidal feature detection rate as a function of star formation activity.
We therefore conclude that the shift in host central color is connected to 
star formation triggered during the merger, rather than stellar mass build-up alone.

This picture of a merger-driven starburst is 
in good agreement with the results of \cite{stierwalt2015}, 
who find that dwarf galaxy pairs with projected separations of $\lesssim 100$ kpc in the
TiNy Titans sample have a significantly higher 
starburst fraction than unpaired dwarfs. Similarly, 
\cite{lelli2014} find that $\sim 80\%$ of a sample of 18 BCDs have a projected
companion within 200 kpc, though they caution that they are not able to verify that
the galaxies will or have collided.
We also extend the results of 
\cite{pustilnik2001}, \cite{ostlin2001}, and \citet{lelli2014}, 
who find high dwarf pair and 
interaction fractions for samples of galaxies known to be BCDs, 
to a larger sample of dwarf
galaxies that span a range of morphologies and star formation rates.

This result can be interpreted in analog to the triggered starbursts observed in 
mergers between more massive galaxies \citep[see, e.g.][]{hopkins2008, ellison2018}. 
However, to our knowledge 
there has not yet been an observational sample of merging dwarf galaxies 
that span a range of star formation activity in the field. 

Though it is widely expected from simulations that mergers will quench
star formation in massive post-merger systems, the nature of the quenching mechanism is 
still unclear. Quenching has been proposed to operate via gas ejection by galaxy-scale 
winds \citep[see, for example,][]{hopkins2008}, 
gas reservoir exhaustion via triggered star formation
\citep[e.g.][]{bekki1998}, or by a highly turbulent post-merger ISM \citep{ellison2018}.
Assuming that the average merger rate increases with redshift 
\citep[see, e.g.][]{fakhouri2010}, the observation of dwarf-dwarf galaxies at $z \lesssim 0.10$ without the presence of a 
significant quiescent population in the field implies that a merger
between two dwarfs is not sufficient for long-term quenching. This apparent
failure to quench may help to constrain both the physical processes
that govern star formation in interacting dwarfs, and the mechanism
behind quenching in higher mass systems.

\subsection{Evidence for a merger pathway for BCD formation}
We furthermore note that the morphologies of the most vigorously starforming galaxies are consistent with that of blue compact dwarfs (BCDs).
BCDs are characterized by the presence of a small number of vigorously star forming
regions that have a spectrum similar to that of HII regions \citep{bekki2008}.
\autoref{starbursts} shows our sample of starbursting 
dwarfs that host tidal features --
though the low surface brightness outskirts of the galaxies 
increase the size of the
galaxy, in shallower imaging only the compact star 
forming core regions would be observable. For example, UM454 (top row, third from left in \autoref{starbursts}) was classified as a ``dwarf HII hotspot''/BCD as early as \cite{salzer1989}.

It has been proposed by \cite{bekki2008} that a possible avenue of 
BCD formation is the merger between two field dwarf galaxies -- 
the increase in the detection fraction for starburst 
galaxies is in good agreement with this formation mechanism. This formation 
mechanism also predicts that older stellar populations will be 
distributed preferentially towards the outskirts of the galaxy, in agreement with
our observations.

\subsection{The origin of non-interacting dwarf starbursts}
A remaining open question is that of 
the chronological link between triggered star formation and tidal feature observability. 
Though the tidal feature detection fraction is significantly higher for high \ha{} EW 
galaxies than for the full sample, we do not see signatures of interaction around all starbursting dwarfs.
Even when including
the results from \cite{stierwalt2015}, who find an increase in the starburst fraction
for dwarfs in pairs, currently detectable dwarf-dwarf 
interactions cannot account for the observed
fraction of dwarfs that are currently in a starburst phase. 

We consider three possibilities to explain the apparently non-interacting starbursting dwarfs:
that the timescale of tidal feature observability is less than the timescale of elevated star formation,
that starbursts can be triggered by mergers more minor than we are able to probe in this study,
and that some fraction of dwarf starbursts are instigated by secular processes. 

The first possibility, that the 
timescale of tidal feature observability is short relative to the 
timescale of elevated star formation rate, 
appears to be unlikely due to the long dynamical times
in the outskirts of dwarf galaxies. 
Though the surface brightness of a tidal stream is expected to 
decay quickly in the first 2 Gyr \citep{johnston2001}, 
we expect the starburst phase to be short 
relative to this timescale \citep{bekki2008}. 

The second possibility, that some starbursts are triggered by mergers more minor than 
those to which we are currently privy, almost certainly accounts 
for some fraction of dwarf 
starbursts. Though the connection between accretion event and starburst trigger is
not easily made for individual objects, we note that 
the nearby starbursting dwarf NGC 4449 hosts a 
disrupting satellite with a stellar mass ratio near
50:1 \citep{rich2012, martinezdelgado2012}.
We are likely only sensitive to recent major mergers, 
where both progenitor galaxies
carry significant gas reservoirs; however, \cite{starkenburg2016} showed that
significant triggered star formation can occur in a 5:1 mass ratio merger, 
even when the incoming 
satellite is a dark halo. 

Finally, secular mechanisms have long been considered 
as an explanation for dwarf starbursts. \cite{elmegreen2012} suggests that the
inspiral of massive gas clumps (on the order of a few percent of the total
galaxy mass) can power starbursts in dwarf galaxies. Similarly, \cite{noguchi2001}
suggests an explanation for BCD formation using a viscosity-driven
mass transport and density threshold for star formation to explain to
compact starforming clumps that characterize BCDs. 

Because we are only sensitive to relatively major mergers between dwarf galaxies,
we are not yet able to distinguish between the picture of a fully merger driven
or partially secular explanation for such starbursts. However, deeper imaging 
will be able to more strongly constrain the contribution of minor mergers to 
the population of starbursting dwarfs. In the absence of this, it may also be possible to detect merging activity with lower mass and/or dark
satellites via kinematic signatures \citep{starkenburg2016b}.

{}

{}

\section{Conclusions and Future Direction}
In this work, we have presented a sample of \nsampauto{} 
isolated dwarf galaxies 
with automatically detected signatures of recent merger
activity selected from \nsampfield{} spectroscopically
confirmed dwarfs in the SDSS and GAMA spectroscopic surveys. The catalog is 
available as a machine-readable table released with this work.

We find that the fraction of galaxies that host detectable tidal features
increases strongly as a function of star formation activity, reaching a tidal 
feature detection of fraction of 15-20\% at the highest \ha{} EW considered 
(${\rm EW}>250$\AA), and that galaxies that host tidal features are systematically
bluer than apparently non-interacting field dwarfs at the same stellar mass. 

Because the origin of irregular LSB structure around dwarf galaxies can be
ambiguous, we construct a \asamp{} of \nsampgold{} galaxies that host unambiguous
merger debris (see the bottom row of \autoref{examples}). 
We find no evidence for a difference in the distribution of 
tidal feature detection fraction between the automated sample and \asamp{} as 
a function of \ha{} equivalent width, host $(g-i)$ color, host stellar mass,
or host redshift. This implies that the automated sample is dominated
by mergers between dwarf galaxies and that contamination (from secular
irregular morphology, spiral arms, etc.) does not change the results 
presented in this study. 

The dependence of tidal feature detection fraction on star formation activity
supports claims that mergers between dwarf galaxies are able to trigger 
starbursts and form blue compact dwarfs \citep{bekki2008}. Observations of 
dwarf galaxies in pairs also show an elevated starburst fraction 
\citep{stierwalt2015}. However, it is not yet clear whether merger-driven
starbursts can explain the entirety of the starbursting dwarf population.

These findings extend observations of star formation in the
dwarf-dwarf merger sequence and show for the first time in a large sample that
merger-driven star formation continues after coalescence, building upon
the evidence for interaction-driven star formation seen in 
dwarf pairs \citep{stierwalt2015, besla2018} and spatially resolved measurements 
of star formation in individual post-merger dwarf systems
\citep{paudel2018, annibali2019}. 

In order to compare these observations of dwarf-dwarf mergers to 
expectations of hierarchical structure formation in $\Lambda$CDM and
to predictions of star formation activity in low-mass systems, it is now
necessary to compare observations and simulations of mergers between
isolated dwarfs in an equivalent manner. Such studies have already been
executed for minor mergers with dark halos 
\citep{starkenburg2015,starkenburg2016,starkenburg2016b}, but expanded 
work concerning the expected population of observable tidal features 
around dwarf galaxies will provide a valuable point of comparison to
simulations.\\

We thank Gurtina Besla, Tjitske Starkenburg, Sarah Pearson, and Kathryn Johnston for constructive conversations regarding this work. We also thank the anonymous referee for their helpful and constructive comments.

This research made use of Astropy, a community-developed core Python package for Astronomy \citep{astropy2018}.

The Hyper Suprime-Cam (HSC) collaboration includes the astronomical communities of Japan and Taiwan, and Princeton University.  The HSC instrumentation and software were developed by the National Astronomical Observatory of Japan (NAOJ), the Kavli Institute for the Physics and Mathematics of the Universe (Kavli IPMU), the University of Tokyo, the High Energy Accelerator Research Organization (KEK), the Academia Sinica Institute for Astronomy and Astrophysics in Taiwan (ASIAA), and Princeton University.  Funding was contributed by the FIRST program from Japanese Cabinet Office, the Ministry of Education, Culture, Sports, Science and Technology (MEXT), the Japan Society for the Promotion of Science (JSPS),  Japan Science and Technology Agency  (JST),  the Toray Science  Foundation, NAOJ, Kavli IPMU, KEK, ASIAA,  and Princeton University.

GAMA is a joint European-Australasian project based around a spectroscopic campaign using the Anglo-Australian Telescope. The GAMA input catalogue is based on data taken from the Sloan Digital Sky Survey and the UKIRT Infrared Deep Sky Survey. Complementary imaging of the GAMA regions is being obtained by a number of independent survey programmes including GALEX MIS, VST KiDS, VISTA VIKING, WISE, Herschel-ATLAS, GMRT and ASKAP providing UV to radio coverage. GAMA is funded by the STFC (UK), the ARC (Australia), the AAO, and the participating institutions. The GAMA website is http://www.gama-survey.org/ .

The Pan-STARRS1 Surveys (PS1) have been made possible through contributions of the Institute for Astronomy, the University of Hawaii, the Pan-STARRS Project Office, the Max-Planck Society and its participating institutes, the Max Planck Institute for Astronomy, Heidelberg and the Max Planck Institute for Extraterrestrial Physics, Garching, The Johns Hopkins University, Durham University, the University of Edinburgh, Queen's University Belfast, the Harvard-Smithsonian Center for Astrophysics, the Las Cumbres Observatory Global Telescope Network Incorporated, the National Central University of Taiwan, the Space Telescope Science Institute, the National Aeronautics and Space Administration under Grant No. NNX08AR22G issued through the Planetary Science Division of the NASA Science Mission Directorate, the National Science Foundation under Grant No. AST-1238877, the University of Maryland, and Eotvos Lorand University (ELTE).

This paper makes use of software developed for the Large Synoptic Survey Telescope. We thank the LSST Project for making their code available as free software at http://dm.lsst.org.

Based in part on data collected at the Subaru Telescope and retrieved from the HSC data archive system, which is operated by the Subaru Telescope and Astronomy Data Center at National Astronomical Observatory of Japan.

Funding for SDSS-III has been provided by the Alfred P. Sloan Foundation, the Participating Institutions, the National Science Foundation, and the U.S. Department of Energy Office of Science. The SDSS-III web site is http://www.sdss3.org/.

SDSS-III is managed by the Astrophysical Research Consortium for the Participating Institutions of the SDSS-III Collaboration including the University of Arizona, the Brazilian Participation Group, Brookhaven National Laboratory, Carnegie Mellon University, University of Florida, the French Participation Group, the German Participation Group, Harvard University, the Instituto de Astrofisica de Canarias, the Michigan State/Notre Dame/JINA Participation Group, Johns Hopkins University, Lawrence Berkeley National Laboratory, Max Planck Institute for Astrophysics, Max Planck Institute for Extraterrestrial Physics, New Mexico State University, New York University, Ohio State University, Pennsylvania State University, University of Portsmouth, Princeton University, the Spanish Participation Group, University of Tokyo, University of Utah, Vanderbilt University, University of Virginia, University of Washington, and Yale University.

\appendix

\section{Isolation and sample selection: robustness tests}\label{appendixB}
\subsection{Isolation Criteria}
It is necessary for this work to create a sample of isolated dwarfs. However, there
exist cases in which different isolation criteria disagree on whether a 
dwarf can be considered isolated. To ensure that our results are not influenced by
these cases, we re-analyze our automated sample with an additional isolation 
criterion.

We adopt the isolation criterion of \cite{paudel2018} because the scope of
the work was to identify mergers between dwarf galaxies. \cite{paudel2018} required
that the dwarf be separated by at least 700 kpc in projection and 700 km s$^{-1}$ from
its nearest massive neighbor, where we adopt a mass cut of 
$M_\star > 4\times 10^{10} M_\odot$ such that the median stellar mass of the 
massive neighbors coincides with that reported by \cite{paudel2018}. This method 
produces a sample of 231 isolated dwarfs, as compared to \nsampauto{} in the
original isolated sample. There are 191 dwarfs that are classified as isolated by
both samples; there are 20 dwarfs that are classified as isolated by the original
criterion and not by the secondary criterion, and 40 dwarfs that are classified as
isolated by the secondary and not the original criterion.

\begin{figure}[ht!]
\center{\includegraphics[width=\linewidth]{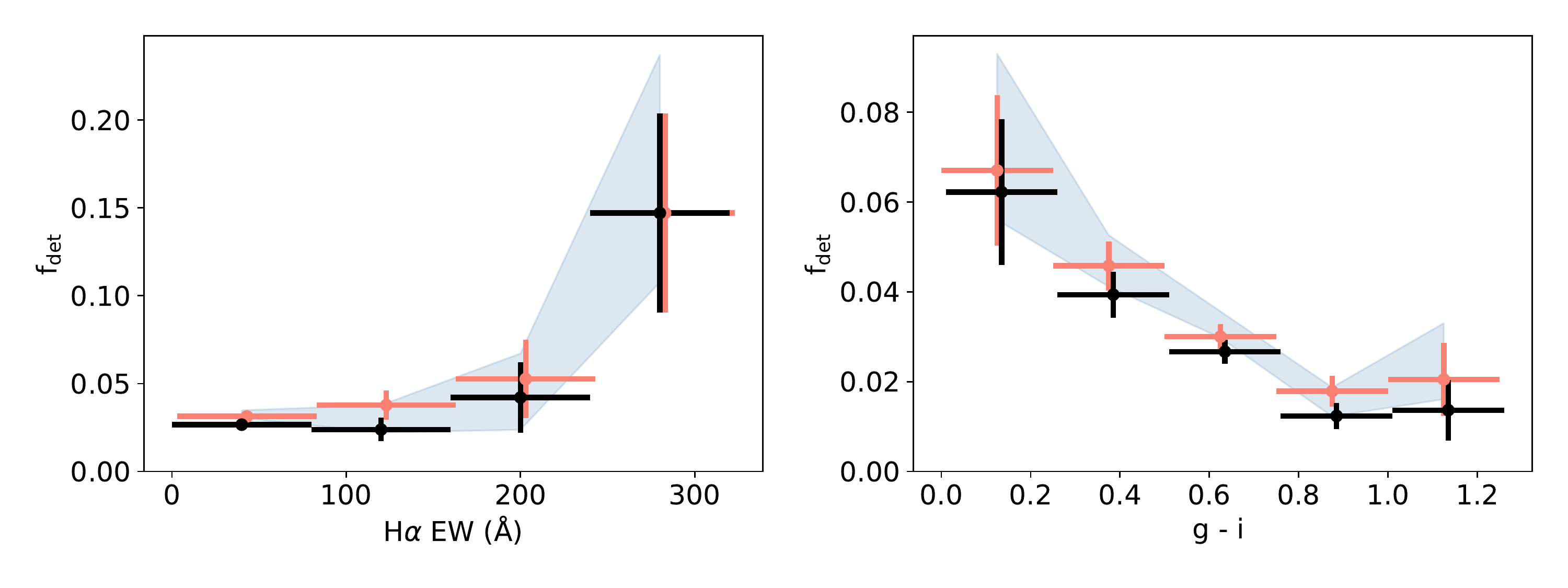}}
\caption{
The same as the analysis of the automated sample in \autoref{sfrcolor_fdet}, 
but where the isolation criterion of \cite{paudel2018} is used to construct
an isolated dwarf sample. The pink errorbars show the detection fraction as a function
of \ha{} equivalent width for dwarf galaxies separated by at least 700 kpc in
projected distance and $\Delta v > 700$ km s$^{-1}$ from their nearest 
massive neighbor. For clarity, these points are offset horizontally. The blue shaded
region shows the results for the full sample (i.e. the same data as
\autoref{sfrcolor_fdet}). The black errorbars show the isolated dwarf sample
created from the intersection of the two isolation criteria (i.e. those galaxies
considered to be isolated by both criteria). All three samples produce statistically
consistent results. 
}
\label{sfrcolor_paudel}
\vspace{20pt}
\end{figure}

We compare this new isolated dwarf sample and the intersection of the two isolated
samples to the original isolated dwarf sample in \autoref{sfrcolor_paudel}. 
The left panel shows the detection fraction as a function of \ha{} equivalent width,
while the right panel shows the same as a function of SDSS (g - i) color. The 
blue shaded regions show the results for the original sample, the pink errorbars for
the new isolated dwarf sample, and the black errorbars for the intersection of the
two samples. Each case produces statistically consistent results, from which we 
conclude that our results are not driven by contamination by non-isolated 
dwarf galaxies.

\subsection{The GAMA and SDSS samples}
Given the heterogeneous characteristics of a sample constructed by
combining dwarf galaxies observed by the GAMA and SDSS campaigns, 
we would like to confirm that we retrieve the same results when considering
the two samples separately.

\begin{figure*}[ht!]
\center{\includegraphics[width=\linewidth]{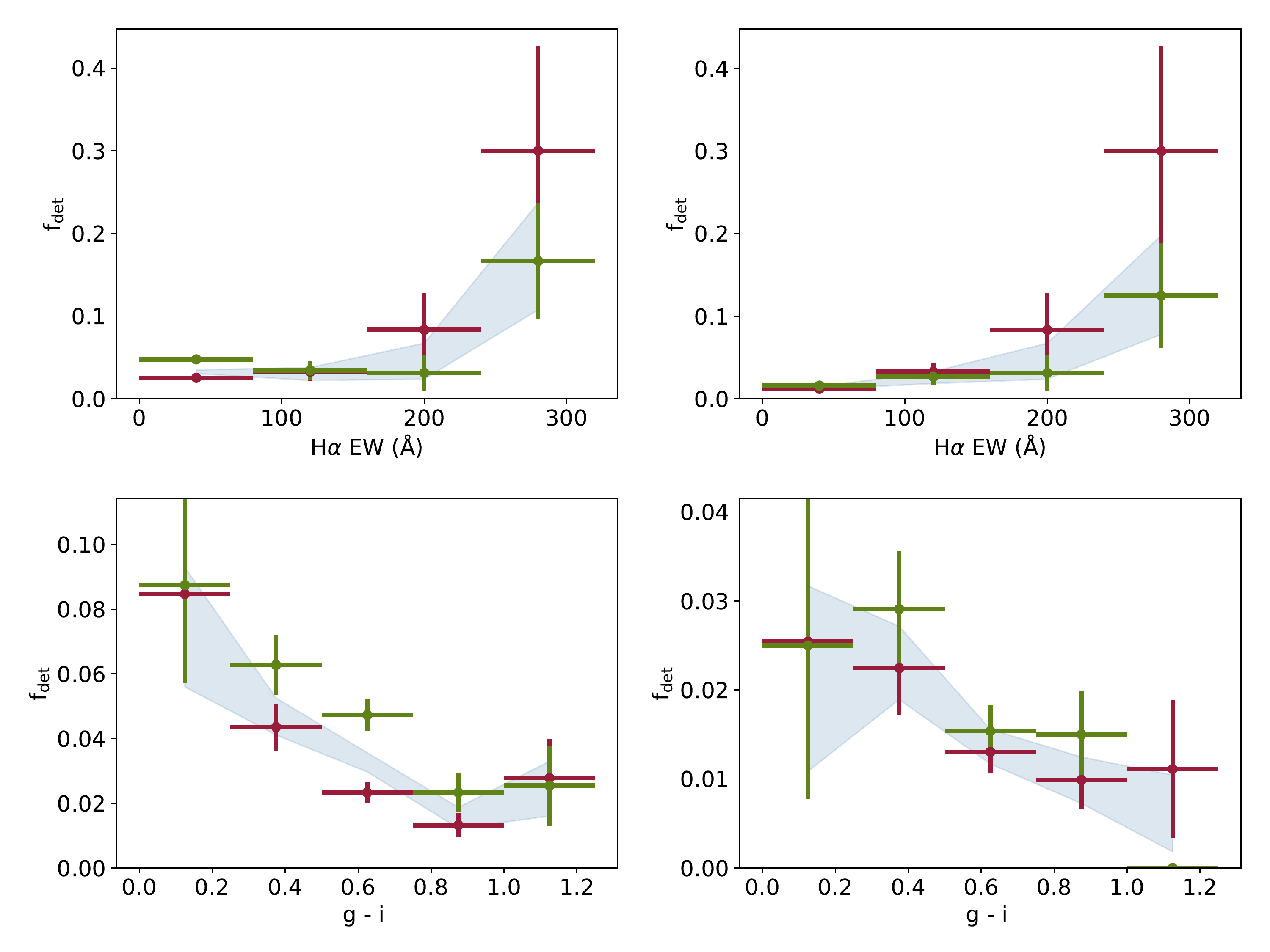}}
\caption{
The same as \autoref{sfrcolor_fdet}, but showing the results from the 
galaxies observed by GAMA (red) and SDSS (green) separately. The blue shaded
region shows the results for the full sample (i.e. the same data as
\autoref{sfrcolor_fdet}). The left column shows the results for the automated
sample, while the right column shows the results for the \asamp{}. In all
cases, the results derived when analyzing the SDSS and GAMA samples
separately are in good agreement with each other and with the results obtained
from the combined sample.
}
\label{subsamp_sfr}
\vspace{20pt}
\end{figure*}

\autoref{subsamp_sfr} is a variation of \autoref{sfrcolor_fdet}, which
presents the tidal feature detection fraction as a function of indications of
star formation activity. In \autoref{subsamp_sfr}, 
the results from the GAMA survey (red) and
the SDSS survey (green) are shown separately, while the grey shaded region
shows the results for the combined sample. The results obtained from the
separated samples are in good agreement with each other and with the results
of the combined sample. We thus conclude that the dependence on star formation
activity that we observe is not driven by the effective selection function
of the parent sample.

\section{Updates to the detection algorithm}\label{appendix}
The updated tidal feature detection algorithm that we employ in this work
differs from the original method in several points. 
First, we now use the starlet wavelet
transform \citep{starck2015} to decompose the input image into 
starlet coefficients that probe a specific spatial scale (previously, we used a
modification of the starlet transform that did not probe as large a range of 
spatial scales). We also now perform multiscale image segmentation to create three dimensional (position-position-wavelet coefficient) 
maps of the LSB debris, remove neighbors and spiral arms in an automated manner, and use flux-weighted nearest neighbor clustering to identify the most probable host for detected LSB features.

We now use the Python package \textsf{sep} \citep{barbary2016},
an implementation of the detection algorithm developed in
\cite{bertin1996}, to detect faint 
features in each difference image individually, 
and construct a final detection map from the features detected at each 
spatial scale.
We detect features at greater than $2\sigma$ significance in the
logarithmic image. This change makes it possible to detect more extended LSB features
that overlap with a large number of foreground and/or background sources; because 
compact sources have power only at small spatial scales, detecting features at each
starlet coefficient separately allows for ``deblending'' as a function of characteristic
spatial extent. To ensure that this change does not characterize the halos of faint sources
as tidal features around bright sources, we require that the flux-weighted center of
each tidal feature be separated from the flux-weighted center of any source by at least
10 pix, set to be approximately twice the PSF FWHM of the worst seeing band ($g_{\rm HSC}$).

We wish to limit contamination from
spiral arms and faint neighbors. To do so, we consider the asymmetry and
clumpiness of the detected candidate tidal feature system.
We follow the historical definition of asymmetry, using

\begin{equation}
\mathcal{A} = \frac{\sqrt{(\sum_{i} f_i - \Tilde{f_i})^2}}{2\sum_i f_i},
\end{equation}
where $f_i$ and $\Tilde{f_i}$ are the flux at position $i$ and the flux at position
$i$ when the image is rotated by 180 degrees, respectively. 
We additionally define the clumpiness of the light 
in the detected LSB debris to be 

\begin{equation}
\mathcal{C} =\sqrt{ \frac{ \langle f_i^2 \rangle}{ \langle f_i \rangle^2 } }.
\end{equation}
In a forthcoming work, we will optimize the values of $\mathcal{A}$ and $\mathcal{C}$
with respect to a visually labeled sample of massive galaxies (Kado-Fong et al. in prep.).
For this work, however,
we choose relatively liberal thresholds that may decrease the purity
of our sample ($\mathcal{A}>0.005$ and
$\mathcal{C} < 20.0$), and instead show that our results are unchanged whether
we consider the automated detection sample or the systems that show
unambiguous merger debris.

Finally, we consider that not all low surface brightness debris in the image cutout is
necessarily associated with our target galaxy. We first remove from the detection all features
less than 10 pixels from the edge of the cutout. Then, we perform a flux-weighted nearest 
neighbor clustering in which each contiguous low surface brightness feature is assigned to 
its most probable host. To do so, we detect the host galaxies via \textsf{sep} in the original 
image and assign a weight inversely proportional to the host flux $w_i = f_i^{-1} / (\sum_j f_j^{-1})$. 

Each contiguous region of the candidate tidal feature system is then assigned to its most
probable host by minimizing the weighted distance between the center of light of the contiguous
region and those of the potential host galaxies. That is to say, the most probable host $j$ minimizes $w_i (\mathbf{x_i} - \mathbf{x^h_j})$.

The full detection 
process is performed independently for the $g_{\rm HSC}$, $r_{\rm HSC}$, and $i_{\rm HSC}$ bands. Only 
detected regions that are present in at least two of bands are retained. This 
decision is in contrast to that of \cite{kadofong2018}, in which only the band with
the best average seeing (the $i_{\rm HSC}$ band) was considered. This change
was made to
remove artifacts via their non-astrophysical colors. 

\section{The Dwarf-dwarf merger catalog}
The galaxies presented in this sample are cataloged and made available via an associated 
machine-readable table. In \autoref{mrt}, we give the positions and classifications of the sources, as well as the stellar mass, color, and \ha{} measurements collated from the GAMA and SDSS catalogs.  We note that we have found three cases in which galaxies are duplicated in the parent sample, either when multiple spectroscopic observations of the same galaxy are marked as distinct objects, or when GAMA and SDSS spectra are assigned to separate objects in the HSC catalog cross-match. We have noted these duplications in the publicly released catalog. However, because we have not inspected every galaxy in the parent sample, and so that it is possible to reproduce the parent sample from the constituent datasets, we do not remove the galaxies from the released catalog. Because these duplications account for a small minority of cases, they do not have a significant impact on the conclusions of this work.

\clearpage
\begin{turnpage}
\begin{table}
\begin{center}
\begin{tabular}{cccccccccccccc}
\toprule
HSC spec-z ID & Spec-z Name &  $\Delta {\rm pos}^1$ &  RA &  Dec &  $z$ &  $\log(\frac{M_\star}{M_\odot})$ & $(g-i)$ &  \ha{} EW &  $f_{H\alpha}$ & $\sigma(f_{H\alpha})$  & automated$^2$ & visual$^3$ & notes$^{5}$ \\\\
{} & {} &  arcsec &  deg &  deg & {}&  {} & mag &  \AA &  $10^{-17}\times$ $\frac{\rm erg}{\rm{s cm}^-2 {\rm \AA}}$  & $10^{-17}\times$ $\frac{\rm erg}{\rm{s cm}^-2 {\rm \AA}}$ & {} &  {}&{} \\\\
\midrule
2895601  &  --$^4$ &   0.06 &  332.76773 &   0.95149 &            0.04 &      9.06 &     0.67 &      23.93 &        73.78 &             1.25 &                False &             False \\
4043445  &   --$^4$ &   0.54 &    18.5433 &  -0.97198 &            0.03 &      8.82 &     0.57 &      30.67 &       80.58 &             2.01 &                False &             False \\
4089749  &  --$^4$ &   0.55 &  351.66805 &   0.59715 &            0.03 &      8.61 &     0.19 &      24.90 &        58.23 &             1.89 &                False &             False \\
4042347  & --$^4$&   0.37 &  333.50968 &  -1.03902 &            0.06 &      9.29 &     0.36 &      73.27 &       267.76 &             3.80 &                False &             False \\
4056352  & --$^4$ &   0.09 &  345.33271 &    -0.654 &            0.08 &      9.49 &     0.75 &      55.90 &       344.49 &             3.74 &                False &             False \\
4056302  & --$^4$&   0.23 &  342.39448 &    0.1745 &            0.02 &      8.39 &     0.55 &      38.56 &       169.98 &             2.82 &                False &             False \\
4074108  &  --$^4$ &   0.19 &  357.97018 &   0.75497 &            0.04 &      9.19 &     0.41 &      77.19 &       504.49 &             4.75 &                False &             False \\
4042561  &  --$^4$ &   0.14 &  344.35628 &  -0.55424 &            0.05 &      7.83 &     1.06 &      14.87 &        86.95 &             3.08 &                False &             False \\
\bottomrule
\bottomrule
\end{tabular}
\caption{This table is published in its entirety in the machine-readable format. A portion is shown here for guidance regarding its form and content. \\
$^1$ The distance between the center of the HSC photometric detection and its nearest spectroscopic counterpart.\\
$^2$ This flag indicates whether the galaxy hosts automatically detected LSB debris.\\
$^3$ This flag indicates whether the LSB debris around the host has been noted as a visually unambiguous product of a dwarf-dwarf merger.\\
$^4$ The ID assigned to the galaxy by the survey from which the redshift is sourced. Due to formatting constraints, we have omitted this column from the text. This information is included in the full machine-readable table associated with this work.\\
$^5$ Notes associated with the source. This is primarily used as a duplicate source flag.
}
\end{center}
\end{table}\label{mrt}
\end{turnpage}
\clearpage

\bibliography{fielddwarfmergers.bib}

\end{document}